\documentclass[aps,pre,preprint, showpacs, superscriptaddress, showkeys]{revtex4}
\usepackage[dvips]{graphicx}
\usepackage{latexsym}
\usepackage{natbib}

\usepackage{color}
\usepackage{amssymb}
\usepackage{multirow}
\usepackage{bigstrut}
\usepackage{subfigure}
\usepackage[percent]{overpic} 
\begin{document}
\title {Fluctuations of a surface relaxation model in interacting
  scale free networks}

\author{M. F. Torres } \affiliation{Instituto de Investigaciones
  F\'isicas de Mar del Plata (IFIMAR)-Physics Department,
  Facultad de Ciencias Exactas y Naturales, Universidad Nacional de
  Mar del Plata-CONICET, Funes 3350, (7600) Mar del Plata, Argentina.}
\author{C.~E.~La~Rocca} \affiliation{Instituto de Investigaciones
  F\'isicas de Mar del Plata (IFIMAR)-Physics Department,
  Facultad de Ciencias Exactas y Naturales, Universidad Nacional de
  Mar del Plata-CONICET, Funes 3350, (7600) Mar del Plata, Argentina.}
\author{L. A. Braunstein} \affiliation{Instituto de Investigaciones
  F\'isicas de Mar del Plata (IFIMAR)-Physics Department,
  Facultad de Ciencias Exactas y Naturales, Universidad Nacional de
  Mar del Plata-CONICET, Funes 3350, (7600) Mar del Plata, Argentina.}
\affiliation{Center for Polymer Studies, Boston University, Boston,
  Massachusetts 02215, USA}


\begin{abstract}

Isolated complex networks have been studied deeply in the last decades due
to the fact that many real systems can be modeled using these types of
structures. However, it is well known that the behavior of a system
not only depends on itself, but usually also depends on the dynamics
of other structures. For this reason, interacting complex networks and
the processes developed on them have been the focus of study of
many researches in the last years.  One of the most studied subjects
in this type of structures is the Synchronization problem, which is
important in a wide variety of processes in real systems. In this
manuscript we study the synchronization of two interacting scale-free
networks, in which each node has $ke$ dependency links with different
nodes in the other network. We map the synchronization problem with an
interface growth, by studying the fluctuations in the steady state of
a scalar field defined in both networks. 

We find that as $ke$ slightly increases from $ke=0$, there is a really
significant decreasing in the fluctuations of the system.  However,
this considerable improvement takes place mainly for small values of
$ke$, when the interaction between networks becomes stronger there is
only a slight change in the fluctuations. We characterize how the
dispersion of the scalar field depends on the internal degree, and we
show that a combination between the decreasing of this dispersion and
the integer nature of our growth model are the responsible for the
behavior of the fluctuations with $ke$.
\end{abstract}


\maketitle

\section{Introduction}

In the last decades the study of complex networks has attracted the
attention of many researchers because many real processes evolve on
these types of structures. In early stages of these studies
researchers were focused on processes that develop on isolated
networks, however, systems, in general, are not completely isolated, but
interacting with other systems instead. These types of interacting
systems, which are a special case of the class called Networks of Networks
(NoN), are composed of internal and external connections. NoN
structures were successfully used to understand epidemic spreading
\cite{Zhao_14_1, Granell, Zuz_15, Zhen_PR, non1}, cascade of failures
\cite{Reis, Bul_01, Gregagos, Val13, non1, Zhen_PR}, diffusion
\cite{Gomez_13, non1, Zhen_PR, Dedom} and synchronization
\cite{Zhen_PR, neuro6, Zhang, Barreto, Matias1}.

 Synchronization phenomena is a relevant subject in many areas, such
 as in neurobiology \cite{neuro6,neuro7,neuro1,neuro2,neuro4}, animal
 behavior \cite{animal1,animal2,pop1}, power-grid networks
 \cite{power1,power2,power3} and so forth. In a relatively recent
 approach, synchronization problems in complex networks are associated
 to the fluctuations of a scalar $h$ defined over the system
 \cite{Korniss,Ana,Cristian,Cristian2,Cristian3,Korniss3,Korniss4,Zoltan1,Zoltan2,Zoltan3,Hunt,Hunt2,Hunt3,Debora,Matias1}.
 This scalar field is a measure of the amount of load that a node has
 to manage. For example, in the problem of queuing networks, the load
 is proportional to the waiting time that a node needs to complete his
 task. The load in a node must be reduced in order to avoid increasing
 the waiting time by distributing efficiently the loads and thus
 improving the synchronization. In this approach the fluctuations are given by
\begin{equation}
W=\sqrt{\left \{\frac{1}{N}\sum_{i=1}^{N}(h_i-\langle h \rangle)^{2}\right \}}\ ,
\end{equation}
where $h_i$ is the load of node $i$, $\langle h \rangle$ is the
average value of the load over the network, $N$ is the system size,
and $\{\}$ is the statistical average.  In the steady state the
fluctuations reach a constant value $W \equiv W_s$, which depends on
the topology of the network. This type of process was studied in
networks with different topologies, but in the last few years many
researches have focused on Scale-Free (SF) networks because they are
obiquous in many real systems. These kinds of networks are
characterized by a degree distribution $P(k)\sim k^{-\lambda}$, where
$P(k)$ is the probability that a node has $k$ internal links and
$\lambda$ is the exponent of the power law distribution. In general,
$\lambda$ takes values between $2.5$ and $3$ in real SF networks.

One of the most studied models of growth interface is the Family model
\cite{Family, Ana, Cristian, Cristian2,
  Cristian3,Debora,Matias1}, which is a surface relaxation model
(SRM).  In this model, at each time-step a node is randomly chosen,
and the node with the lowest amount of load or `height' between the
selected node and its neighbors increases its load in one unit. In
 isolated SF networks with exponent $\lambda < 3$ it was
found that the
dependence of $W_s$ with the system size $N$ has a crossover between two
different behaviors at a characteristic size $N_0$: for $N<N_0$, $W_s
\sim log N$ , and $W_s$ $\sim$ constant for $N>N_0$
\cite{Debora}. Thus in the last regime the system is scalable,
i.e. increasing the system size does not affect the fluctuations.  In
a more recent work \cite{Matias1} the authors studied the SRM in two
interacting SF networks, in which a fraction $q$ ($0\leq q \leq 1$) of
nodes in each network is connected one by one through bidirectional
external links, allowing diffusion from one network to another.
They found that the synchronization improves as $q$ increases and reaches an
 improvement of $40\%$ for $q=1$.  In real systems however,
nodes can have more than one external connection with nodes in the
other networks, which implies a stronger interaction between the
systems. This strong interaction may affect the processes that develop
on structures of this type. In this work we are interested in
understanding how the strong interaction between networks affects the
synchronization of the system. For this purpose we study the SRM model
in two SF networks in which each node has $ke$ external connections. 
In this study we only use stochastic simulations due to the fact that
 the heterogeneity of the SF networks and the lack of geometrical
 direction makes difficult any theoretical approach \cite{Cristian}.

\section{Model}

We build two SF networks $A_i$ ($i=1,2$) using the Molloy--Reed
Algorithm \cite{Molloy}, avoiding multiple and self connections, and
we use a minimum degree $k_{min}=2$ to ensure that each network has a
single component \cite{Cohen}. In both networks every node $j$, with
$j=1,..,N$, has $k_j^i$ internal connections with nodes in the same
network and $ke$ external connections with nodes in the other
network. By simplicity, we consider the same number of external
connections for all nodes. We denote by $v_j^i$ and $b_j^i$ the set of
internal and external neighboring nodes respectively of node $j$ from network $A_i$. We chose as initial condition all the scalar fields
$h_j^i$ randomly distributed in the interval $[0,1]$.

At each time step a node $j$ in one of the networks $A_i$, with
$i=1,2$, is randomly chosen and receives a load unit. Then:

1) The load diffuses to the node $m$, which is the one with the
smaller load in the set $\{j,v_j^i\}$.  We denote this process as
the first internal diffusion.
  
\begin{figure}[h]
  \begin{center}
    \includegraphics[width=0.5\textwidth]{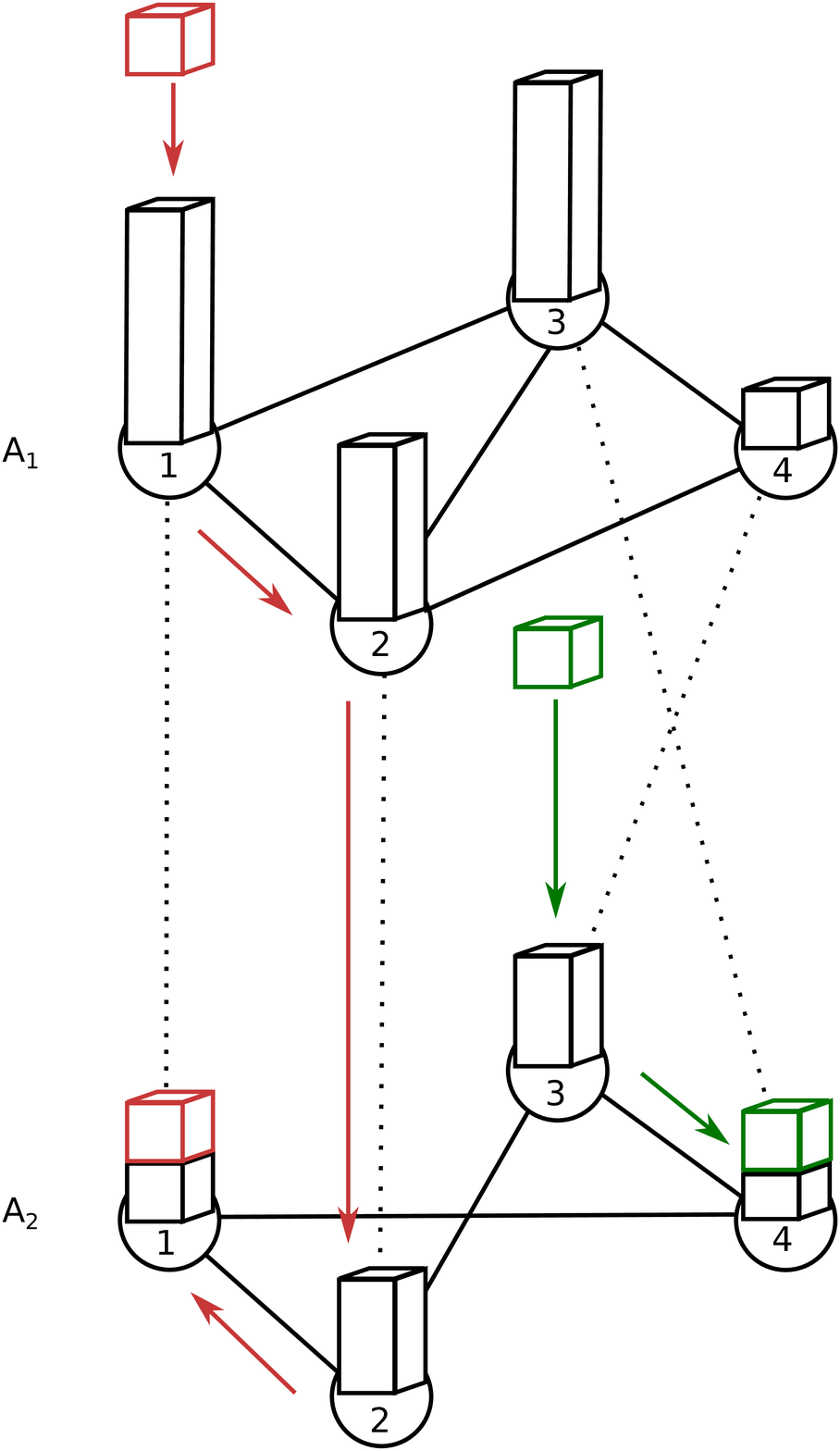}
    \caption{Schematic of the rules of our model for two interacting
      networks. The circles represent the nodes and the columns the
      amount of load or height in each node. Full lines denote the
      internal connections and the dot-lines the external ones. The networks have $N=4$ and $ke=1$. A load unit is
      dropped in a node of one of the networks (colored cubes) and
      diffuses between nodes following the rules (colored arrows),
      until it is finally deposited.}
    \label{fig.1}
  \end{center}
\end{figure}

2) If $h_m^i$ is smaller than all the heights in the set $b_m^i$, then
the load is deposited in $m$ and $h_m^i=h_m^i+1$. (color green in Fig
\ref{fig.1}).  Otherwise the load diffuses to the node with the
smaller height in the set $\{b_m^i\}$ . We denote this process as external
diffusion.

3) If an external diffusion takes place, step $1)$ is repeated and,
after a second internal diffusion, the load is deposited in a node $n$
in the network $A_l$ with $l\neq i$ (color red in Fig
\ref{fig.1}). Then $h_n^l=h_n^l+1$.

\section{Results and Discussions}

\begin{figure}[h]
  \begin{center}
    \includegraphics[width=0.5\textwidth]{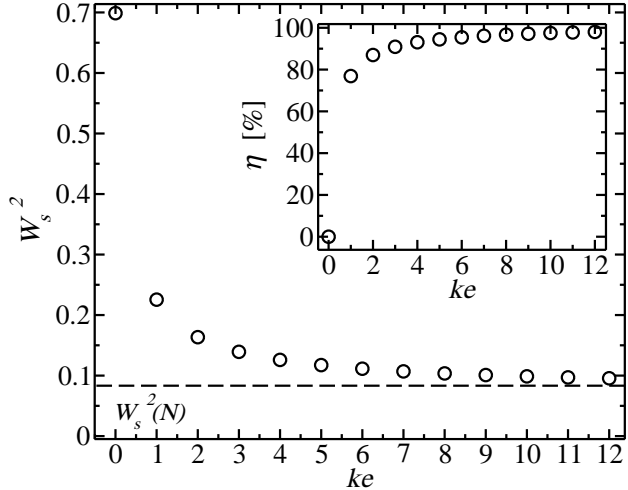}
    \caption{Square fluctuations of the scalar field $W^2_s$ as a
      function of the external connections $ke$ ($\circ$). The dashed
      line corresponds to the case $ke=N$. In the inset we plot the
      percentage of the maximum improvement that the synchronization
      can achieve compared to the isolated network $\eta$ as a
      function of $ke$. For small values of $ke$, $\eta$ almost
      approaches its optimal value.}
    \label{fig.2}
  \end{center}
\end{figure}

For the simulations we build two SF networks with the same exponents
$\lambda=2.6$ and sizes $N=3 \times 10^5$ to ensure that the system is
in the scalable regime \cite{Debora}. As the two networks used here
have the same exponent $\lambda$ and same system size $N$, the
fluctuations $W^i_s$ on each network will be in average the same, thus
by simplicity we drop the index $i$. In Fig. \ref{fig.2} we plot the
square fluctuations in the steady state of each network $W^2_s$ as a
function of the external connection parameter $ke$. We can observe
that the synchronization of the system improves as $ke$ increases and
that the fluctuations converge asymptotically to the optimal value
$W_s^2(N)$, which corresponds to the case $ke=N$.  The reduction in
the fluctuations when more external connections are added is due to
the fact that the overloaded nodes in one network have the possibility
to diffuse their excess of load to nodes that possesess lower levels of
load in the other network. This external diffusion allows to
synchronize nodes that have increasingly similar amounts of load. It
is important to notice that for high interacting networks with $k_e=N
$, the value $W_s^2(N)$ is independent of the exponent $\lambda$ of
the degree distribution because in this case the interaction between
networks, represented by the external connections, dominates the
dynamics and the saturation value of the synchronization. Thus in this
case of strong interaction the topology of the isolated
networks represented by internal connections plays a minor role. More precisely, the nodes have more probability to diffuse to nodes in
the other network because the number of external connections is much
higher than the number of internal ones.

In order to visualize the relative improvement of the synchronization
compared to the isolated network we compute the percentage of the
maximum optimization that the system achieves for $ke > 0$,
$\eta=(W_s^2(0)-W_s^2(ke))/(W_s^2(0)-W_s^2(N))\times 100\%$.  In the
inset of Fig. \ref{fig.2} we plot $\eta$ as function of $ke$. From the
plot we can observe that we do not need to have very high values of
$ke$ to get a good improvement, since $\eta$ approaches fast to the
optimal case ($ke=N$). As $ke$ increases more connections are added in
order to obtain some improvement in the synchronization, and this is
very expensive compared to the resulting profit. For example,
increasing $ke$ from zero to five implies adding $5\;N$ external
connections and $\eta$ is about $94.5\%$, when we increase $ke$ from
$5$ to $10$ we duplicate the external connections between networks
and obtain only a $3\%$ of improvement over the previous case. This is
a high cost to pay to only obtain a slight improvement in the
synchronization of the system.

\begin{figure}[h]
\begin{center}
\includegraphics[width=0.5\textwidth]{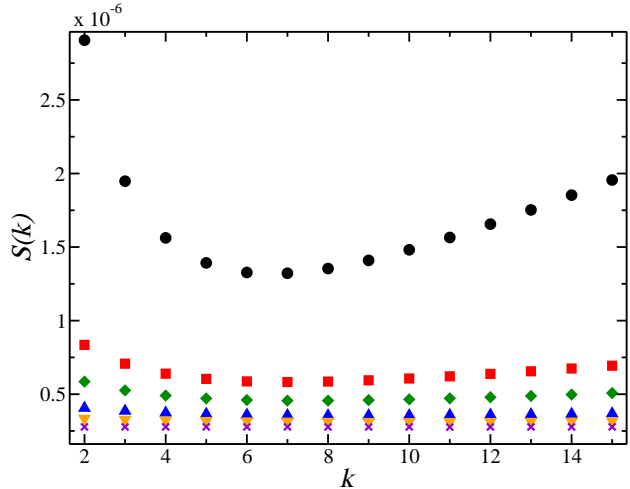}
\caption{The average contribution $S(k)$ that a node with degree $k$
  makes to the fluctuations as a function of $k$, for $ke=0$ (black
  $\circ$), $ke=1$ (red $\Box$), $ke=2$ (green $\Diamond$), $ke=5$
  (blue $\bigtriangleup$), $ke=10$ (orange $\bigtriangledown$), and
  $ke=N$ (violet $\times$). The values of $S(k)$ shown in the vertical
  axis are multiplied by $10^{-6}$. The curves exhibit a minimum
  around $k=6$ for all $ke$. For $ke=N $, $S(k)$ is a constant which
  does not depend on the topology of the networks.}
\label{fig.3}
\end{center}
\end{figure}

To understand the role of the internal connections of the nodes in the
fluctuations, we define the average contribution $S(k)$ that nodes
with degree $k$ made to $W^2_s$, where $S(k)$ is given by
\[ S(k)=\frac{1}{N}\sum_{j=1 \;/ k_j=k}^N\frac{(h_j - \langle h \rangle)^2}{N\;P(k)}\ .\]
It is straightforward to show that
$W^2_s=\sum_{k=k_{min}}^{k_{max}}N\;P(k)\;S(k)$. In Fig. \ref{fig.3}
we plot $S(k)$ as a function of the internal degree $k$ for different
values of $ke$.  From this plot we can see that as $k$ increases
$S(k)$ decreases, reaching a minimum around $k=6$. This behavior can
be explained as follows: as $k$ increases the nodes have more
neighboring nodes to which they can send their excess of load or from
which they can receive load, approaching their height to $\langle h
\rangle$. However around $k=6$, the nodes have too many neighbors and
start to act as load sinks, and thus high degree nodes become the most
loaded of the system.  As $ke$ increases $S(k)$ decreases for all the
values of $k$ and as a consequence the synchronization improves, and
in addition $S(k)$ has a minor dependence on $k$. This implies that
the amount of load of all the nodes becomes similar and hence the
external connections dominate the dynamics of the growth model over
the internal topology of the network. Thus $S(k)$ has the same value
regardless of the internal degree of a node when $ke=N$. In
Fig.~\ref{fig.3} we show only the values of $S(k)$ for nodes with
$k<15$ because the total contribution from higher degree nodes to the
fluctuations is far smaller.

\begin{figure}[h]
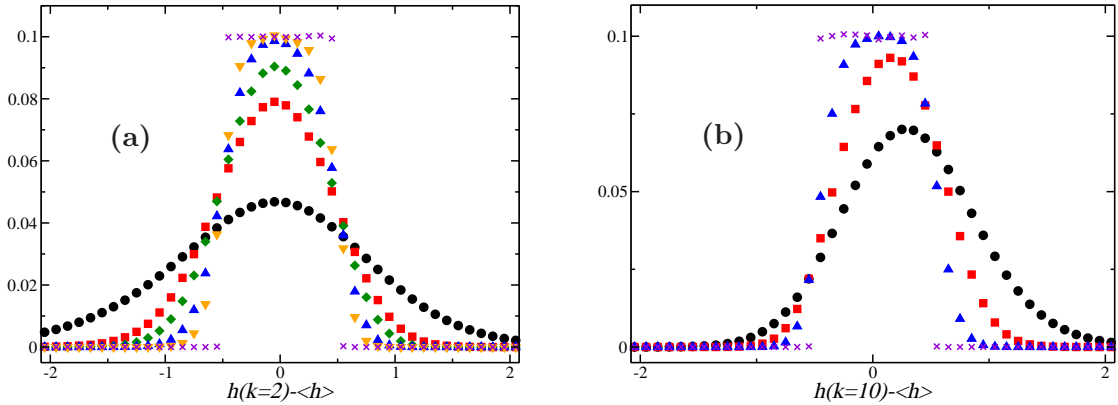
 
\vspace{1cm}
\begin{center}
  \begin{overpic}[scale=0.25]{Fig3.eps}
    \put(20,50){\bf{(a)}}
\end{overpic}\hspace{1cm}\vspace{1cm}
  \begin{overpic}[scale=0.25]{Fig4.eps}
    \put(20,50){{\bf{(b)}}}
  \end{overpic}
\vspace{-1cm}
\caption{(a) Distribution of load of the nodes around the mean value for
  nodes with (a) degree $k=2$ for $ke=0$ (black
  $\circ$), $ke=1$ (red $\Box$), $ke=2$ (green $\Diamond$), $ke=5$
  (blue $\bigtriangleup$), $ke=10$ (orange $\bigtriangledown$) and
  $ke=N$ (violet $\times$). (b) degree $k=10$ for $ke=0$ (black
  $\circ$), $ke=1$ (red $\Box$), $ke=5$ (blue $\bigtriangleup$) and
  $ke=N$ (violet $\times$). }
\label{fig.4}
\end{center}
\end{figure}

 In order to explain why the main contributions to the fluctuations
 are reduced when the interaction between networks increases, we study
 the distribution of load of nodes with internal connectivity $k$
 around the main value $h(k) - \langle h\rangle$, where $h(k)$ is the
 average amount of load of nodes with degree $k$. In Fig. \ref{fig.4} we plot
 the distributions of $h(k) - \langle h\rangle$, for $k=2$ and for
 $k=10$ for different values of $ke$. From Fig.~\ref{fig.4}(a), we can
 see that as the number of external connections increases for $k_e >
 0$, the dispersion of these distributions is reduced. For example,
 when we increase $ke$ from $0$ to $1$, the dispersion is reduced from
 $0.934$ to $0.501$. Also these distributions have nodes with levels
 of load above and below the mean value, and with an average load
 very close to $\langle h \rangle$, regardless of the value of $ke$.
 These observations, which are also seen for $k<6$ --not shown here--
 confirm that the main reason of the reduction of the fluctuations is
 that nodes with small degree and load above the main value send
 their excess of load to nodes with also small degree and load below
 the main value through the external connections of the system. Other
 values of dispersion of these distributions around the main value are
 displayed in Table \ref{TabDisp}.

 From Fig.~\ref{fig.4}(b) we can see that the
distribution of $h(k=10) - \langle h\rangle$ for $ke=0$ is not
centered in $0$. This is because, as explained above, nodes with $k>6$
act as load sinks and usually become over saturated in the growing
process. As $ke$ increases not only the width of the distribution is
reduced but also it is centered close to $0$. Thus all the load
distributions for different values of $k$ get closer as $ke$
increases. Adding external connections decreases the dispersion of the
distributions, however due to the discrete nature of the rules of our
growth model and the initial conditions, the reduction has a limit at
which the distribution tends to a rectangular shape as can be seen
in Fig.~\ref{fig.4}(a) and Fig.~\ref{fig.4}(b) for $ke \to N$. Also
from Table \ref{TabDisp} we can see that the values of the
dispersion of the load distribution for $k=2$ are closer to the values
for $k=10$ as $ke$ increases. All the results shown in this table are
qualitatively the same for other values of $\lambda$ in the interval
$2 < \lambda \leq 3$. 

It is worthwhile to mention that we expect that in NoN
composed by more than two networks $W^2_s$ will be slightly smaller
than in the case of two interacting networks, because the system will
be able to perform more relaxation steps between networks.

\begin{table}
\caption{Dispersion of the distributions of load of the nodes with
  internal degree $k$ around $\langle h \rangle$ for different values
  of $ke$. Notice that the dispersion is $\sqrt{N\;S(k)}$. }
\label{TabDisp}
\begin{center}
\begin{tabular}{|c||c|c|c|c|c|c|}
\hline

 $ke$ & 0        & 1         &  2      & 5  & 10 & N         \bigstrut\\ \hline
 $k=2$ & 0.934         & 0.501         &  0.419      & 0.349  & 0.317 & 0.289    \bigstrut \\
$k=10$ & 0.667           & 0.427         & 0.373        &0.326     &0.306 & 0.289     \bigstrut \\ \hline

\end{tabular}

\end{center}
\end{table}

\section{Conclusions}
We studied the synchronization of a system of two interacting SF
networks for a simple surface relaxation model, in which the nodes of
each network are connected with $ke$ nodes of the other network. We
found that as $ke$ increases the synchronization of the system
improves, reaching an optimal value for $ke=N$. We also found that the
addition of a small number of $ke$ connections results in a similar
improvement compared to the case $ke=N$. This is important because an
increase in the value of $ke$ requires to add a large number of
external connections to the system, which is very expensive from an
economic point of view. We explain the reduction of the fluctuations
in each network when $ke$ increases, by the decrease of the
contributions from nodes with small degree, which is a consequence of the external
connections that reduce the dispersion of load of these nodes around the
mean value.

In future works we will explore another type of external connection,
such as external connections taken from a distribution.

\section{Acknowledgments}

CEL and LAB want to thank UNMdP, FONCyT, Pict 0429/2013, Pict
1407/2014, CONICET, PIP 00443/2014 for financial support. MFT
acknowledges CONICET, PIP 00629/2014 for financial support.

\end{document}